\documentstyle[preprint,revtex]{aps}
\tightenlines
\begin{document}
\draft

\begin{title}
Pressure Dependence of the Elastic Moduli in \\
Aluminum Rich Al--Li Compounds
\end{title}

\author{Michael J. Mehl\cite{MJM}}

\begin{instit}
Complex Systems Theory Branch,\\
Naval Research Laboratory, \\
Washington, D.C.  20375-5000

{\em

Email: {\tt mehl@irrmasg6.epfl.ch} \\
After 30 Sept 1992: {\tt mehl@dave.nrl.navy.mil} \\
Preprint Number: cond-mat/9208001 \\
Submitted to Phys. Rev. B1:~~13 July 1992\\

}

\end{instit}

\receipt{}

\begin{abstract}
I have carried out numerical first principles calculations of the
pressure dependence of the elastic moduli for several ordered
structures in the Aluminum-Lithium system, specifically FCC Al, FCC
and BCC Li, L1$_2$ Al$_3$Li, and an ordered FCC Al$_7$Li supercell.
The calculations were performed using the full potential linear
augmented plane wave method (LAPW) to calculate the total energy as
a function of strain, after which the data was fit to a polynomial
function of the strain to determine the modulus.  A procedure for
estimating the errors in this process is also given.  The predicted
equilibrium lattice parameters are slightly smaller than found
experimentally, consistent with other LDA calculations.  The
computed elastic moduli are within approximately 10\% of the
experimentally measured moduli, provided the calculations are
carried out at the experimental lattice constant.  The LDA
equilibrium shear modulus $C_{11}-C_{12}$ increases from 59.3~GPa in
Al, to 76.0~GPa in Al$_7$Li, to 106.2~GPa in Al$_3$Li.  The modulus
$C_{44}$ increases from 38.4~GPa in Al to 46.1~GPa in Al$_7$Li, then
falls to 40.7~GPa in Al$_3$Li.  All of the calculated elastic moduli
increase with pressure with the exception of BCC Li, which becomes
elastically unstable at about 2~GPa, where $C_{11}-C_{12}$ vanishes.
\end{abstract}

\section{Introduction}
\label{sec:intro}
The ductility, light weight, and high strength of Aluminum make it
the best choice for many areas of construction.  It can, however, be
improved.  The addition of small amounts of Lithium will greatly
increase the strength of the alloy with little change in the
ductility \cite{Muller86}.  In particular, Young's modulus increases
{}from 70~GPa in pure Aluminum to 94~GPa when the atomic concentration
of Lithium reaches 32\% \cite{Nobel82}.  The experimental value of
$C_{11}-C_{12}$ in the metastable L1$_2$ phase of Al$_3$Li is some
50\% larger than it is in pure Al \cite{Axon48}.

In order to study the effect of Lithium additions on Aluminum, I
used the Linear Augmented Plane Wave (LAPW) method \cite{Andersen75}
to calculate the equation of state and the pressure dependence of
the elastic moduli of FCC Aluminum, FCC and BCC Lithium, the L1$_2$
phase of Al$_3$Li, and an ordered FCC Al$_7$Li superlattice.  The
results of these calculations may be used as a database for fitting
approximate methods, such as the Connolly and Williams (CW) method
\cite{Burton92}, the generalized Ising Hamiltonian method
\cite{Wei92}, or the embedded atom method \cite{Rifkin92}.  Although
calculations of this type have been carried out in the Al-Li system
to determine the phase diagram \cite{Sluiter90,Podloucky88}, the
elastic moduli were not computed at that time.

The results of these calculations are interesting in their own right
because they show the behavior of Al$_x$Li over a wide range of
pressures.  They also demonstrate the accuracy which can be achieved
by the LAPW method even when the difference in energies between the
different structures is very small (less than 1 mRy).

The outline of this paper is as follows.  In Sec.~\ref{sec:elastic}
I summarize the methods used for calculating elastic moduli.
Sec.~\ref{sec:fitting} shows how to estimate the errors in the
calculations.  The results (C$_{ij}$ versus pressure) are presented
in Sec.~\ref{sec:results}.  Sec.~\ref{sec:summary} contains a
summary of the results.

\section{Calculation of the Elastic Moduli}
\label{sec:elastic}

A previous paper \cite{Mehl90} described the method used to
determine the elastic moduli of intermetallic alloys from total
energy calculations.  I will present a brief review of the method
here.

The total energy calculations are carried out using a full potential
version \cite{Wei85a} of the Linear Augmented Plane Wave Method
\cite{Andersen75}.  I will refer to this as ``the LAPW method''.
The transformation from the many-body problem to a single-particle
picture is obtained via the Hedin-Lundqvist parameterization
\cite{Hedin71} of the Local Density Approximation \cite{Kohn65} to
the density functional theory \cite{Hohenberg64}.  The LAPW method
treats the core states (here the Al 1s, 2s, and 2p states, as well
as the Li 1s states) fully relativistically, and treats the
conduction bands in the semi-relativistic approximation
\cite{Koelling77}.  The lattice is divided into muffin-tin spheres
and an interstitial region.  Inside the muffin-tins the basis
functions are expanded into spherical harmonics up to order $l = 8$.
In the interstitial region the basis functions are expanded into
plane waves.  I experimented with the basis set cutoff, and found
that at the equilibrium volume approximately 40 wave functions/atom
were sufficient in the sense that the predicted elastic moduli did
not change when I increased the number of basis functions.  The
single particle potential is also also expanded into spherical
harmonics and plane waves, however the spherical harmonic expansion
inside the muffin-tins is only carried out up to terms of order $l =
4$.  The Brillouin-zone integrations were performed using the
Monkhorst and Pack special {\bf k}-point prescription
\cite{Monkhorst76}, with modifications to properly treat the special
{\bf k}-points in reduced symmetry lattices \cite{Mehl90}.

The elastic moduli of a cubic crystal may be divided into two
classes, the bulk modulus $B = (C_{11} + 2 C_{12})/3$, and the two
shear moduli, $C_{11} - C_{12}$ and $C_{44}$.  The bulk modulus is
related to the curvature of $E(V)$,
\begin{equation}
B(V) = - V P'(V) = V E''(V) \; , \label{bulk}
\end{equation}
where $V$ is the volume of the unit cell, $E(V)$ is the energy/unit
cell at volume $V$, and $P(V) = -E'(V)$ is the pressure required to
keep the cell at volume V.  Since the calculations only provide a
set of energies $E(V_i)$ for a limited number of volumes $V_i$, the
second derivative $E''(V)$ must be approximated.  Here I begin by
making a least squares fit of the computed energies to the form
proposed by Birch
\cite{Birch78}:
\begin{eqnarray}
E(V) & = &E_o + \frac98 B_oV_o [(\frac{V_o}V)^{2/3}-1]^2 +
\nonumber\\
& & \frac9{16} B_o(B_o'-4)V_o[(\frac{V_o}V)^{2/3}-1]^3 +
\sum_{n=4}^N \gamma_n [(\frac{V_o}V)^{2/3}-1]^n\;. \label{birn}
\end{eqnarray}
where $E_o$, $V_o$, $B_o$, and $B_o'$ are, respectively, the
equilibrium energy, volume, bulk modulus, and pressure derivative of
the bulk modulus, while N is the order of the fit.  For a second
order fit ($N = 2$) it is obvious that $B_o' = 4$.  Experimentally,
$B_o'$ is usually between 3 and 5.  The bulk modulus can then be
obtained by analytic differentiation of (\ref{birn}).

The shear moduli require knowledge of the derivative of the energy
as a function of a lattice strain \cite{Kittel86}.  In the case of a
cubic lattice, it is possible to choose this strain so that the
volume of the unit cell is preserved.  The strain can also be chosen
so that the energy is an even function of the strain, whence an
expansion of the energy in powers of the strain contains no odd
powers.  Thus for the calculation of the modulus $C_{11}-C_{12}$ I
used the volume-conserving orthorhombic strain tensor,
\begin{equation}
\stackrel{\leftrightarrow}{\varepsilon} =
\left(\matrix{\delta&0&0\cr 0&-\delta&0\cr
0&0&\delta^2/(1-\delta^2)}\right) \; .
\label{ortho}
\end{equation}
Application of this strain changes the total energy from its
unstrained value to
\begin{equation}
E(\delta) = E(-\delta) = E(0) + (C_{11}-C_{12}) V \delta^2 +
O[\delta^4] \; ,
\label{orthoE}
\end{equation}
where V is the volume of the unit cell and $E(0)$ is the energy of
the unstrained lattice at volume V.  For the elastic modulus
$C_{44}$, I used the volume-conserving monoclinic strain tensor
\begin{equation}
\stackrel{\leftrightarrow}{\varepsilon} =
\left(\matrix{0&\frac12\delta&0\cr \frac12\delta&0&0\cr
0&0&\delta^2/(4-\delta^2)}\right) \; .
\label{mono}
\end{equation}
which changes the total energy to
\begin{equation}
E(\delta) = E(-\delta) = E(0) + \frac12 C_{44} V \delta^2 +
O[\delta^4] \; .
\label{monoE}
\end{equation}
Note that there is no pressure or stress term \cite{Alouani91} in
either (\ref{orthoE}) or (\ref{monoE}) since the strains
(\ref{ortho}) and (\ref{mono}) are constructed so that $\Delta V =
0$.

The strains (\ref{ortho}) and (\ref{mono}) can be used for any cubic
lattice.  In the general case, the internal parameters of the
lattice must be chosen to minimize the total energy of the strained
structure.  Fortunately, the lattices discussed here fall into a
restricted subset of the cubic lattices, where all of the atoms sit
at inversion sites, even under the reduced symmetry caused by the
strains (\ref{ortho}) and (\ref{mono}).  For lattices in this class,
the atoms will remain on the inversion sites for infinitesimal
strains $\delta$.  The sites may become unstable at some finite
strain, but even then the force on the atoms at the inversion sites
will be zero.  Since the elastic modulus is only concerned with the
limit $\delta \rightarrow 0 $, it is unnecessary to relax the
internal parameters in the strained lattice.  This enormously
simplifies the calculation.  Furthermore, the elimination of the
linear term ensures that $\delta = 0$ will be a extremal point, as
required by symmetry.

The use of the strains (\ref{ortho}) and (\ref{mono}) reduces the
symmetry of the problem compared to the tetragonal and trigonal
distortions which are sometimes used \cite{Mehl90}.  The lower
symmetry means that more {\bf k}-points are generated, however this
is compensated by the fact that (\ref{orthoE}) and (\ref{monoE}) are
even functions of the strain $\delta$, and so we need only one-half
as many calculations.

Having outlined the methods used to calculate the $C_{ij}$, it must
be noted that many compounds, including some of those in the Al-Li
system, cannot be grown in single crystal form, hence the individual
elastic moduli cannot be measured experimentally.  In these cases
experiments can only determine the isotropic bulk $B$ and shear $G$
moduli of polycrystalline aggregates of small crystallites.  In the
case of the cubic systems discussed here the calculated $C_{ij}$ can
be used to determine $B$ exactly and to place rather strict bounds
on $G$.  General bounds on $B$ and $G$ were originally determined by
Reuss \cite{Reuss29} and Voight \cite{Voight28}.  Later, Hashin and
Shtrikman found improved bounds specific to cubic materials
\cite{Hashin62}.  We will use the later bounds here. For isotropic
polycrystalline aggregates of cubic crystallites, the bulk modulus
is given exactly by
\begin{equation}
B = \frac13 (C_{11} + 2 C_{12}) \label{agbulk} \; ,
\end{equation}
just as for a cubic crystal \cite{Schreiber73}.  In true isotropic
materials, the shear modulus is related to the elastic moduli by
\begin{equation}
G^I = C_{44}^I = (C_{11}^I - C_{12}^I)/2 \; ,
\end{equation}
but in real crystals the anisotropy constant
\begin{equation}
A = \frac{2 C_{44}}{C_{11}-C_{12}} \label{aniso}
\end{equation}
is not unity.  In this case, we can only bound the shear modulus of
the aggregate.  Hashin and Shtrikman \cite{Hashin62} found that for
cubic crystals these bounds are given by
\begin{equation}
G_1 = G_1^* + \frac{3 (G_2^* - G_1^*)}{5 - 4 \beta_1 (G_2^* -
G_1^*)}
\end{equation}
and
\begin{equation}
G_2 = G_2^* + \frac{2 (G_1^* - G_2^*)}{5 - 6 \beta_2 (G_1^* -
G_2^*)} \; ,
\end{equation}
where
\begin{equation}
G_1^* = \frac12 (C_{11} - C_{12}) , \;\; G_2^* = C_{44}
\end{equation}
and
\begin{equation}
\beta_1 = - \frac{3(B+2G_1^*)}{5G_1^*(3B+4G_1^*)} , \;\;
\beta_2 = - \frac{3(B+2G_2^*)}{5G_2^*(3B+4G_2^*)} \; .
\end{equation}
The Shtrikman bound $G_S$ is designated as the smaller of $G_1$ and
$G_2$, while the Hashin bound $G_H$ is the larger.  Note that in the
limit of an isotropic lattice ($A = 1$, or $G_1^* = G_2^*$) we get
$G_S = G_H = G$.

In the Al-Li system these bounds are so tight that the difference
$|G_H-G_S|$ is on the order of the uncertainty in the calculation of
the individual $C_{ij}$.  Thus I will use
\begin{equation}
G = \frac12 (G_S+G_H) \label{agshear}
\end{equation}
as the definition of the shear modulus.  The estimated error in G is
on the order of the estimated error in the individual $C_{ij}$.
Associated with the bulk and shear modulus are Young's modulus
\begin{equation}
E = \frac{9 B G}{3 B + G} \label{Young}
\end{equation}
and Poisson's ratio
\begin{equation}
\sigma = \frac12 (1 - \frac{E}{3B}) \label{Poisson}
\end{equation}
for isotropic crystals.  Since these quantities are often of
interest I will list them in the results.

\section{Estimating Errors}
\label{sec:fitting}
Using equations (\ref{orthoE}) and (\ref{monoE}), the elastic moduli
$C_{11}-C_{12}$ and $C_{44}$ are simply related to the second
derivative of $E(\delta)$ at zero strain.  If $\delta'$ is an
infinitesimal strain, then, since we have eliminated terms linear in
$\delta$,
\begin{equation}
\alpha = \lim_{\delta' \rightarrow 0} \frac{E(\delta') - E(0)}{V
\delta'^2} \; ,
\label{infinitesimal}
\end{equation}
where $\alpha$ is the linear combination of the $C_{ij}$ associated
with the strain $\delta$.  In principle, we could pick a small,
finite strain $\delta$, calculate $E(\delta)$ and $E(0)$, and use
(\ref{infinitesimal}) to estimate the elastic modulus.  In practice,
however, numerical errors, which cause fluctuations in the computed
$E(\delta)$, make the calculated value of $\alpha$ change rapidly
and unpredictably with $\delta$ unless great care is taken to ensure
convergence in both {\bf k}-points and basis set size.

To avoid this difficulty, it is better to chose a set of M points
$\delta_i$ and fit the resulting $E(\delta)$ to a polynomial of the form
\begin{equation}
\epsilon_N[\{f_n\};\delta] = \sum_{n=0}^N f_n \delta^{2n} \; .
\label{poly}
\end{equation}
The polynomial (\ref{poly}) contains no odd powers in $\delta$
because these terms have been eliminated from the expansions
(\ref{orthoE}) and (\ref{monoE}).  The quantity $f_1/V$ plays the
role of $\alpha$ in Eq.~(\ref{infinitesimal}).  Note that $f_1$ may
still be very sensitive to the choice of $N$ and the $\delta_i$.  We
need a reliable procedure for determining the best value of $f_1$
{}from the available data.  This procedure should determine the order
$2N$ of the polynomial expansion (\ref{poly}), and also give some
estimate of the possible values of $f_1$ which would give a nearly
equivalent fit.  To do this analysis, I have chosen to follow a
method outlined in the {\em Numerical Recipes} book \cite{Press86}.

The calculation of an elastic modulus $C_{ij}$ from total energy
calculations begins by choosing a set of $M$ points $\delta_i, ~\{i
= 1, 2, ... , M\}$.  Typically I choose $M = 5$, with $\delta_1 = 0$
and $\delta_5 \approx 0.05$.  If the $\delta_i$\ are equally spaced,
then for the Aluminum-bearing compounds discussed here $\Delta E_i =
E(\delta_{i+1}) - E(\delta_i)$ is on the order of 1 mRy.  Smaller
values of $\Delta E$ may cause excess numerical noise to introduce
large uncertainties in the analysis.

To fit (\ref{poly}) there must be an energy $E_i$ and an error
estimate $\sigma_i$ for each $\delta_i$.  I estimated $E_i$ and
$\sigma_i$ by performing LAPW total energy calculations on several
{\bf k}-point meshes.  For FCC Li and Al I used meshes corresponding
to 32000, 42592, and 55296 points in the full Brillouin zone.
Symmetry reduced the actual number of points to no more than 7200 in
the irreducible part of the Brillouin zone.  For structures with
larger unit cells and hence smaller Brillouin zones, correspondingly
smaller numbers of {\bf k}-points were used.  In retrospect, this
large number of {\bf k}-points may be excessive, but $\Delta
E(\delta)$ in Li is rather small (the energy change between $\delta
= 0$ and $\delta = 0.1$ is about 1 mRy), so high accuracy was
needed.  In any case, since the secular equation was small (about 40
basis functions), these calculations could be done quickly even for
large numbers of {\bf k}-points.  The energy $E_i$ was calculated by
taking an average of the computed $E(\delta_i)$ over the different
meshes, weighted by the number of points in the mesh.  The
uncertainty in the calculation, $\sigma_i$, was assumed to be the
{\bf k}-point weighted standard deviation about this average.  This
method of determining $\sigma_i$ does not include errors associated
with other variables in the calculation, including basis set size
and the Fast Fourier Transform mesh size \cite{Wei85b}.  However, it
should give at least an order of magnitude estimate of the size of
the error in the calculation.

For a given N, the parameters $f_n$ in (\ref{poly}) are chosen by a
least squares fitting procedure, minimizing the quantity
\begin{equation}
\chi^2(N) = \sum_{i=1}^M \left(\frac{E_i -
\epsilon_N[\{f_n\};\delta_1]}{\sigma_1}\right)^2 \; .
\label{chi}
\end{equation}
If the numerical errors are normally distributed, the probability
that an error less than or equal to $\chi^2(N)$ can occur by chance
is
\begin{equation}
q(N) = {\int_{\chi^2(N)/2}^{\infty} t^{(M-N-3)/2} e^{-t} \;
dt}/{\Gamma((M-N-1)/2)} \; . \label{qeq}
\end{equation}
For a given compound and lattice constant, the best N is the one
which maximizes $q(N)$.  Typically, the best q was on the order of
0.5, indicating that our error estimates $\sigma_i$ were probably
too large \cite{Press86}.  (Usual values of $\sigma$ are on the
order of 0.03 mRy for these calculations.)  Maximum values of q
ranged from 0.05 to 0.95.

As an example of this process, Fig.~\ref{a750o} plots the energy of
L1$_2$ Al$_3$Li as a function of the orthorhombic strain
(\ref{ortho}) for the fixed volume associated with the cubic lattice
constant $a = 7.50$~a.u. ($r_s = 2.93$~a.u.).  The $N=1$ and $N=2$
fits are also shown.  Obviously the $N=1$ fit ($q = 0.002$) is worse
than the $N=2$ ($q = 0.99$) fit.  On the other hand, as seen in
Fig.~\ref{a750m}, $N=1$ ($q=0.68$) and $N=2$ ($q=0.74$) fit the
monoclinic strain energy (\ref{monoE}) equally well.

Once the order N of the fit (\ref{poly}) has been determined, the
accuracy of $f_1$, the parameter related to the elastic modulus, can
be established.  Assume that the numerical errors in the calculation
are normally distributed.  Then there is a 68.3\% probability that
the correct $f_1$ is within the set of all possible values of $f_1$
which change $\chi^2(N)$ by less than 2.3.  The bounds I present for
the elastic moduli are those selected by this formula.

The equilibrium parameters and confidence limits derived from the
Birch fit ($E_o$, $V_o$, $B_o$, and $B_o'$) are obtained by similar
methods using the functional form (\ref{birn}) and first-principles
data $E(V_i)$.

I have written out the above procedure to show how an estimate of
the errors in these calculations can be obtained and so that other
researchers will be able to obtain the same elastic moduli and
estimated uncertainties from the same set of data.  The accuracy of
these uncertainties is obviously open to question.  The errors here
are probably not normally distributed.  In fact, some systematic
errors, such as the change in the energy with respect to basis set
size, are not included in the analysis.  To conclude, then, these
error estimates only serve as a guide to the uncertainty in these
calculations.

\section{Results}
\label{sec:results}

The results are presented as a series of tables and figures.  I
first used the LAPW method to calculate the total energy versus
volume for FCC Al, FCC Li, L1$_2$ Al$_3$Li, and ordered Al$_7$Li in
an FCC supercell, as diagrammed in Fig.~\ref{al7li}.  For
completeness, calculations were also performed for BCC Li.  The data
{}from these calculations is shown in Fig.~\ref{engeos}, where the
energy/atom is plotted versus the Wigner-Seitz radius.  The data was
fit to the Birch equation (\ref{birn}), as described in
Sec.~\ref{sec:fitting}.  In all cases the $N=3$ Birch fits had the
best ``q'' value (\ref{qeq}).  Table~\ref{equtab} lists the
calculated equilibrium lattice constants, bulk moduli, and pressure
derivatives for the $N=3$ Birch fits.  I also list a theoretical
binding energy, calculated by finding the difference between the
LAPW total/energy atom and the self-consistent total energy of a
spherically averaged atom using the Hedin-Lundqvist \cite{Hedin71}
approximation to the LDA and the semi-relativistic approximation
\cite{Koelling77} for the ``valence'' orbitals.  The computed results
are in good agreement with the calculations of Guo {\em et al.}
\cite{Guo90a,Guo90b}, which I also present here.

The available experimental data for the lattice constants
\cite{Donohue74,Pearson67} and bulk modulus
\cite{Brandes83,Simmons71} are also listed in Table~\ref{equtab}.
The predicted lattice constants are all smaller than experiment.
The error is largest in BCC Lithium, where the lattice constant is
3\% smaller than experiment.  The Al and Al$_3$Li lattice constants
are within about 1\% of experiment.  While some of this is due to
the neglect of zero point motion and thermal expansion, much of the
error, especially in Lithium, can be attributed to problems with the
LDA \cite{Perdew92}.

The equation of state, $P(V)$, for these curves can be determined by
differentiating the fit (\ref{birn}).  The volume dependence of the
bulk modulus, $B(V)$, can be found by applying the formula
(\ref{bulk}) to (\ref{birn}).  Since $P(V)$ is a monotonic function
over the range of volumes shown here it is trivial to numerically
invert this function and thus calculate $B(P)$, the pressure
dependence of the bulk modulus.  Fig.~\ref{bulkp} plots B(P) for the
systems studied in the paper.

The bulk moduli are in reasonably good agreement with experiment.
The predicted Lithium modulus is about 26\% too large, while the Al
modulus is 8\% too large.  The latter error is typical of other
calculations \cite{Mehl90}.  Since the LDA underestimates the
equilibrium lattice constant, it is not surprising that it
overestimates the bulk modulus.  If we use the Birch fit
(\ref{birn}) to evaluate the bulk modulus (\ref{bulk}) at the
experimental lattice constants, we get bulk moduli of 10.6~GPa for
BCC Lithium and 66.7~GPa for Aluminum.  These values are both 12\%
lower than the respective experimental values, overcorrecting the
experimental values.  It is not clear whether this remaining error
is due solely to the LDA or if contributions to the bulk modulus
{}from the thermal and zero-point motion should be included.

The shear moduli were somewhat more difficult to obtain, since the
small variation in the total energy as a function of the shear
$\delta$ required a large {\bf k}-point mesh.  The uncertainties in
these moduli are usually larger than the uncertainties in $B_o$.
Fig.~\ref{c11-c12} and Fig.~\ref{c44} show $C_{11}-C_{12}$ and
$C_{44}$, respectively, as a function of pressure.  The error bars
show the estimated uncertainty in the moduli (see
Sec.~\ref{sec:fitting}).  The moduli were actually calculated at
fixed volume.  The pressure was calculated from the Birch fit
(\ref{birn}) using $P(V) = - E'(V)$.

As one can see from the figures, the pressure dependence of the
elastic moduli can be fit by a straight line, at least over the
range $-2$~GPa to $12$~GPa.  The lines shown in the figures are
determined by a least squares fit to the computed elastic moduli,
using the estimated uncertainty as a weighting factor.  The
intercepts of these lines with the $P=0$ axis are listed in
Table~\ref{shearmod}, while the slopes are in Table~\ref{shearprs}.
The estimated errors in the fit are calculated in the manner of
Sec.~\ref{sec:fitting}.

Table~\ref{shearmod} also lists the available experimental data for
Aluminum \cite{Simmons71}, and the Al$_3$Li calculations of Guo {\em
et al.} \cite{Guo91}.  There is a slight volume discrepancy between
the computed Aluminum moduli and the experimental numbers, so it is
not surprising that the calculated moduli are larger than
experiment.  Evaluating the moduli at the experimental volume gives
$C_{11}-C_{12} = 48.7$~GPa and $C_{44} = 30.1$~GPa, both within 8\%
of experiment.  Combining these numbers with the bulk modulus data
shown above, it seems that the LDA can predict the elastic moduli to
within about 10\%, if the experimental volume is used.

Except for $C_{12}$, the computed Al$_3$Li moduli are somewhat
smaller than the calculations in Ref.~\cite{Guo91} or the
experimental results reported there.  Those calculated moduli were
presumably determined at the experimental volume, so the true
discrepancy is somewhat larger than the 42\% disagreement in
$C_{44}$ shown in the table.  However, both sets of calculations
agree that there is a large increase in $C_{11}-C_{12}$ with
increasing lithium content.

I next used the zero pressure moduli from Table~\ref{shearmod} to
calculate the Hashin-Shtrikman bounds on the shear modulus $G$ for a
polycrystalline aggregate.  Notice that, except for Li, the bounds
on $G$ are smaller than the uncertainties in the original elastic
moduli.  Thus the actual uncertainties in $G$ are approximately the
same as the uncertainties in Table~\ref{shearmod}.
Table~\ref{anitab} collects the information which can be derived for
the aggregate crystals, including the bulk modulus (\ref{agbulk}),
the shear modulus (\ref{agshear}), Young's modulus (\ref{Young}),
and Poisson's ratio (\ref{Poisson}).  The Young's modulus for
several Al-Li compounds has been measured experimentally
\cite{Nobel82}, and is also presented in Table~\ref{shearmod}.  Not
surprisingly, the predicted Young's modulus is some 23\% larger than
the experimental number.  Evaluating the modulus at the experimental
parameter gives $E = 103~GPa$, within 13\% of experiment.

Although not part of the major thrust of this paper, there are
several features of interest in the Lithium data.  Calculations show
that FCC Li is 0.2~mRy more stable than BCC Li.  Experimentally, BCC
Lithium may transform to the FCC state at $T = 78K$
\cite{Donohue74}, although the actual ground state of Lithium is
apparently a hexagonal close packed structure.  The calculations are
also in agreement with earlier works \cite{Nobel92}, which agree
that the FCC-BCC energy difference is about 0.2~mRy.

The behavior of Lithium under pressure is also worthy of note.
Fig.~\ref{c11-c12} shows that BCC Lithium becomes elastically
unstable at about 2~GPa, when the modulus $C_{11}-C_{12}$ vanishes.
This is consistent with the ground state of Lithium being a
close-packed phase \cite{Donohue74}.  However, the Birch fits to the
equation of state show a pressure induced phase transition from FCC
Lithium {\em to} BCC Lithium near 0.6~GPa.  Since the energy
difference between the two phases is only 0.2~mRy, it is obvious
that thermal and zero point motion, which are neglected here, must
be included to obtain an understanding of the behavior of Lithium.

\section{Summary}
\label{sec:summary}

As shown above, addition of Lithium into FCC Aluminum increases both
of the shear moduli.  $C_{11}-C_{12}$ is 28\% larger in Al$_7$Li
than in pure Al, and 79\% larger in Al$_3$Li than in Al.  For
$C_{44}$ the increases are 20\% and 6\%, respectively.  Conversely,
the bulk modulus decreases slightly with increasing Lithium
concentration.  Since $C_{11}-C_{12}$ and $C_{44}$ increase by
nearly the same ratio in going from Al to Al$_7$Li, the anisotropy
factor $A$ (\ref{aniso}) is nearly unchanged, while it is much
smaller in Al$_3$Li than it is in either Al or Al$_7$Li.  This
increase in anisotropy was also found by Guo {\em et al.}
\cite{Guo91}, who attributed it to increasing anisotropy of the
chemical bonding with the addition of Lithium.  From the results
presented here, it is apparent that the anisotropy ratio does not
change appreciably for small amounts of Lithium, increasing only
when there are large numbers of Lithium atoms on next-nearest
neighbor sites (see Fig.~\ref{al7li}).  However, Poisson's ratio
(\ref{Poisson}) does decrease from 0.32 in Al to 0.26 in Al$_7$Li,
indicating that some anisotropy is present even at this Li
concentration.  It would be interesting to study the behavior of the
anisotropy at smaller Lithium concentrations by continuing the
elastic moduli calculations for ordered supercells with chemical
composition Al$_{15}$Li or Al$_{26}$Li.  Unfortunately, while simple
total energy calculations for these structures are possible on
present computers, in this case the not all of the Aluminum atoms
are at inversion sites, so relaxation of these atoms around the
Lithium ``impurities'' must be included \cite{Mehl91}.  It would
require a large number of total energy calculations to obtain the
energies of the relaxed structures needed for the calculation of the
$C_{ij}$ in these systems.

The elastic moduli all change linearly with pressure in the region
studied (-2 to 12~GPa).  There is some variation in the slope of
these lines, but generally the slope is between 3 and 5.  The
exceptions are $C_{11}'-C_{12}'$ and $C_{44}'$ in the Lithium
structures, which are all less than 1, and $C_{11}'$ in the Aluminum
compounds, where the slope is about 7.

In conclusion, I have calculated the pressure dependence of the
elastic moduli for several ordered Al-Li lattices, including error
estimates in the calculation of the elastic moduli.  The
computations show that the addition of Lithium strongly increases
the shear moduli and Young's moduli (by about 20\%) over Aluminum.
The decrease in Poisson's ratio indicates that the Lithium
introduces anisotropic chemical bonding.

\acknowledgments

I would like to thank D. A. Papaconstantopoulous, B. M. Klein, and
D. Singh for helpful comments and encouragement.  Special thanks are
due to A.  Gonis, who suggested this investigation, and to R.
Podloucky, who lead me to several references.  Parts of these
calculations were done at the {\em Institut Romand de Recherche
Num\'{e}rique en Physique de Mat\'{e}riaux} (IRRMA), Lausanne,
Switzerland.  Parts of this work are supported by the Office of
Naval Research, United States Department of Defense.

\figure{Energy as a function of the square of the orthorhombic
strain (\ref{ortho}), used to determine $C_{11}-C_{12}$ in L1$_2$
Al$_3$Li.  The unstrained cubic lattice constant is $a = 7.50$~a.u.
($r_s = 2.93$~a.u.).  The error bars, representing the estimated
error $\sigma_i$, defined in the text, are not shown as they are
smaller than the heights of the symbols.  The solid line is the
$N=1$ fit (\ref{poly}), the dashed line is the $N=2$ fit.
\label{a750o}}

\figure{Energy as a function of the square of the monoclinic strain
(\ref{mono}), used to determine $C_{44}$ in L1$_2$ Al$_3$Li.  The
unstrained cubic lattice constant is $a = 7.50$~a.u. ($r_s = 2.93
$~a.u.).  The error bars on the data points, which represent the
errors $\sigma_i$ defined in the text, are omitted as they are
almost the same height as the symbols.  The solid line is the $N=1$
fit (\ref{poly}), the dashed line is the $N=2$ fit. \label{a750m}}

\figure{The FCC based structures used in this paper.  The FCC
lattice sites are represented by open, shaded, and filled circles.
When all sites contain the same type of atom, this represents a
simple FCC lattice, with the cube volume equal to four times the
volume of the unit cell.  If the filled sites are occupied by Al
atoms and the open and shaded sites by Li atoms, this represents
Al$_3$Li in the L1$_2$ structure, with the cube shown being the
entire unit cell.  Finally, if the shaded and filled sites are
occupied by Al atoms and the open sites by Li atoms, this represents
a part of the FCC supercell Al$_7$Li lattice.  The eight cubes which
touch at the lower left-hand corner of the lattice form a cube
enclosing four times the volume of the Al$_7$Li unit cell.
\label{al7li}}

\figure{Computed energy/atom (in Rydbergs) versus the Wigner-Seitz
atomic radius $r_s$ (in atomic units) for compounds in the Al-Li
system.  The energy is relative to the atomic energies calculated
for spherically averaged atoms.  (See the text.)  The corresponding
$N=3$ Birch fits (\ref{birn}) are plotted as continuous lines.  In
the upper part of the graph, the solid line and the ``+'' represent
BCC Li, the dashed line and the ``$\diamond$'' FCC Li.  In the lower
part of the graph, from the bottom up the curves are FCC Al (solid
line and ``$\Box$''), FCC Al$_7$Li (dashed line and ``$\times$''),
and L1$_2$ Al$_3$Li (dotted line and ``$\bigtriangleup$'').  The
error bars representing the $\sigma_i$ would be smaller than the
size of the markers and so are not shown.
\label{engeos}}

\figure{Computed bulk modulus versus pressure.  All units are GPa.
Since this is computed from (\ref{bulk}), using the Birch fit
(\ref{birn}) in Fig.~\ref{engeos}, no data points are shown.  In the
lower part of the graph the solid line represents BCC Li, the dashed
line FCC Li.  In the upper part of the graph the solid, dashed and
dotted lines represent FCC Al, FCC Al$_7$Li, and L1$_2$ Al$_3$Li,
respectively.  Over this pressure range $B(P)$ can be reasonably
approximated by a straight line.
\label{bulkp}}

\figure{Shear modulus $C_{11}-C_{12}$ versus pressure.  All units
are GPa.  In the lower part of the graph, the solid line and the
``+'' represent BCC Li, the dashed line and the ``$\diamond$'' FCC
Li.  In the upper part of the graph, from the bottom up the curves
are FCC Al (solid line and ``$\Box$''), FCC Al$_7$Li (dashed line
and ``$\times$''), and L1$_2$ Al$_3$Li (dotted line and
``$\bigtriangleup$'').  The calculation of the error bars is
described in Sec.~\ref{sec:fitting}.  The straight lines are least
squares fits to the calculated pressure dependence of
$C_{11}-C_{12}$, weighted by the estimated error. \label{c11-c12}}

\figure{Shear modulus $C_{44}$ versus pressure.  All units are in
GPa.  In the lower part of the graph, the solid line and the ``+''
represent BCC Li, the dashed line and the ``$\diamond$'' FCC Li.  In
the upper part of the graph, from the bottom up the curves are FCC
Al (solid line and ``$\Box$''), L1$_2$ Al$_3$Li (dotted line and
``$\bigtriangleup$''), and FCC Al$_7$Li (dashed line and
``$\times$'').  The calculation of the error bars is
described in Sec.~\ref{sec:fitting}.  The straight lines are least
squares fits to the calculated $C_{44}(P)$, weighted by the
estimated error.  Note that the Al$_7$Li and Al$_3$Li curves are
reversed compared to Fig.~\ref{c11-c12}. \label{c44}}

\narrowtext
\begin{table}
\caption{Equilibrium constants for cubic structures in the Al-Li
system, using the $N=3$ Birch fit (\ref{birn}).  The Wigner-Seitz
radius $r_s$ is in atomic units, the bulk modulus $B_o$ in GPa, and
the cohesive energies $E_c$ in Rydbergs.}
\begin{tabular}{llcccc}
&System&$r_s$&$B_o$&$B_o'$&$E_c$\\
\tableline
BCC &Li$^{\rm a}$ &\dec 3.13 &\dec 15.1$\pm.2$ &\dec 3.1$\pm0.5$
&\dec -0.1511 \\
BCC &Li$^{\rm b}$ &\dec 3.13 &\dec 15.0 & &\dec
-0.1245 \\
BCC &Li&\dec 3.24$^{\rm c}$ &\dec 12.0$^{\rm d}$ & & \\
FCC &Li$^{\rm a}$ &\dec 3.16 &\dec 15.0$\pm.1$ &\dec 3.4$\pm0.2$
&\dec -0.1513 \\
FCC &Li$^{\rm b}$ &\dec 3.12 &\dec 14.0 & &\dec
-0.1250 \\
FCC &Li &\dec 3.24$^{\rm c}$ & & & \\
FCC &Al$^{\rm a}$ &\dec 2.95 &\dec 82.4$\pm.9$ &\dec 4.8$\pm0.1$
&\dec -0.308 \\
FCC &Al$^{\rm e}$ &\dec 2.95 &\dec 82.0 & &\dec -0.295 \\
FCC &Al&\dec 2.99$^{\rm c}$ &\dec 76.3$^{\rm f}$ & & \\
FCC &Al$_7$Li$^{\rm a}$ &\dec 2.94 &\dec 75.6$\pm.6$ &\dec
5.0$\pm0.3$ &\dec -0.292 \\
FCC &Al$_7$Li$^{\rm e}$ &\dec 2.94 &\dec 74.0 & &\dec -0.283 \\
L1$_2$ &Al$_3$Li$^{\rm a}$ &\dec 2.93 &\dec 69.1$\pm.5$ &\dec
4.7$\pm1.2$ &\dec -0.277 \\
L1$_2$ &Al$_3$Li$^{\rm e}$ &\dec 2.94 &\dec 72.0 & &\dec -0.272 \\
L1$_2$ &Al$_3$Li &\dec 2.95$^{\rm g}$ & & & \\
\end{tabular}
\label{equtab}
\tablenotes{$^{\rm a}$Calculated, this work.}
\tablenotes{$^{\rm b}$Calculated, Ref. \cite{Guo90b}.}
\tablenotes{$^{\rm c}$Experiment, Ref. \cite{Donohue74}.}
\tablenotes{$^{\rm d}$Experiment, Ref. \cite{Brandes83}.}
\tablenotes{$^{\rm e}$Calculated, Ref. \cite{Guo90a}.}
\tablenotes{$^{\rm f}$Experiment, Ref. \cite{Simmons71}.}
\tablenotes{$^{\rm g}$Experiment, Ref. \cite{Pearson67}.}
\end{table}

\mediumtext
\begin{table}
\caption{Equilibrium elastic moduli for cubic structures in the
Al-Li system.  The Wigner-Seitz radius $r_s$ is in atomic units, and
represents the equilibrium lattice constant for computations and the
room-temperature lattice constant for experimental measurements.
All moduli are in GPa.  The last column lists the anisotropy ratio,
$A$ (\ref{aniso}).}
\begin{tabular}{llccccccc}
&System&$r_s$&$B$&$C_{11}-C_{12}$&$C_{11}$&$C_{12}$&$C_{44}$&A \\
\tableline
BCC &Li$^{\rm a}$ &\dec 3.13 &\dec 15.1$\pm.2$ &\dec 0.8$\pm0.1$
&\dec 15.6$\pm0.3$ &\dec 14.8$\pm0.3$ &\dec 11.1$\pm0.2$ &\dec 27.8
\\
FCC &Li$^{\rm a}$ &\dec 3.16 &\dec 15.0$\pm.1$ &\dec 3.9$\pm0.6$
&\dec 17.6$\pm0.5$ &\dec 13.6$\pm0.3$ &\dec 9.1$\pm0.2$ &\dec 4.6 \\
FCC &Al$^{\rm a}$ &\dec 2.95 &\dec 82.4$\pm.9$ &\dec 59.3$\pm1.0$
&\dec 121.9$\pm1.6$ &\dec 62.7$\pm1.3$ &\dec 38.4$\pm3.0$ &\dec 1.3
\\
FCC &Al$^{\rm b}$ &\dec 2.99 &\dec 76.3 &\dec 46.0 &\dec 107.0 &\dec
61.0 &\dec 28.0 &\dec 1.2 \\
FCC &Al$_7$Li$^{\rm a}$ &\dec 2.94 &\dec 75.6$\pm.6$ &\dec
76.0$\pm3.0$ &\dec 126.2$\pm2.6$ &\dec 50.3$\pm1.6$ &\dec
46.1$\pm2.7$ &\dec 1.2 \\
L1$_2$ &Al$_3$Li$^{\rm a}$ &\dec 2.93 &\dec 69.1$\pm.5$ &\dec
106.2$\pm3.8$ &\dec 139.8$\pm3.0$ &\dec 33.7$\pm1.7$ &\dec
40.7$\pm2.2$ &\dec 0.8 \\
L1$_2$ &Al$_3$Li$^{\rm c}$ & &\dec 72.3 &\dec 128.6 &\dec 158.0
&\dec 29.4 &\dec 57.7 &\dec 0.9 \\
\end{tabular}
\label{shearmod}
\tablenotes{$^{\rm a}$ This work.}
\tablenotes{$^{\rm b}$ Experiment, Ref. \cite{Simmons71}.}
\tablenotes{$^{\rm c}$ Calculations, Ref. \cite{Guo91}.}
\end{table}

\begin{table}
\caption{Equilibrium pressure derivatives of the elastic moduli for cubic
structures in the Al-Li system.  The Wigner-Seitz radius $r_s$ is in
atomic units.}
\begin{tabular}{llcccccc}
&System&$r_s$&$B'$&$C_{11}'-C_{12}'$&$C_{11}'$&$C_{12}'$&$C_{44}'$
\\
\tableline
BCC &Li &\dec 3.13 &\dec 3.1$\pm0.5$ &\dec -0.3$\pm0.03$ &\dec
2.9$\pm0.6$ &\dec 3.2$\pm0.5$ &\dec 0.7$\pm0.1$ \\
FCC &Li &\dec 3.16 &\dec 3.4$\pm0.2$ &\dec 0.4$\pm0.4$ &\dec
3.6$\pm0.5$ &\dec 3.2$\pm0.3$ &\dec 0.5$\pm0.1$ \\
FCC &Al &\dec 2.99 &\dec 4.8$\pm0.1$ &\dec 3.3$\pm0.3$ &\dec
7.0$\pm0.4$ &\dec 3.7$\pm0.2$ &\dec 2.6$\pm1.0$ \\
FCC &Al$_7$Li &\dec 2.94 &\dec 5.0$\pm0.3$ &\dec 2.2$\pm0.7$ &\dec
6.5$\pm0.7$ &\dec 4.3$\pm0.5$ &\dec 3.1$\pm0.6$ \\
L1$_2$ &Al$_3$Li &\dec 2.93 &\dec 4.7$\pm0.2$ &\dec 4.4$\pm1.2$
&\dec 7.6$\pm0.9$ &\dec 3.2$\pm0.5$ &\dec 2.5$\pm0.7$ \\
\end{tabular}
\label{shearprs}
\end{table}

\begin{table}
\caption{Derived properties of aggregate crystals.  The moduli $B$,
$G_S$, $G_H$, $G$, and $E$ are in GPa.  All calculations are done
using the LDA equilibrium moduli in Table~\ref{shearmod}.}
\begin{tabular}{llcccccc}
& System & $B$ & $G_S$ & $G_H$ & $G$ & $E$ & $\sigma$ \\
\tableline
BCC & Li &\dec 15.1 &\dec 1.6 &\dec 5.1 &\dec 3.3 &\dec 9.3 &\dec
0.40 \\
FCC & Li &\dec 15.0 &\dec 4.6 &\dec 5.4 &\dec 5.0 &\dec 13.5 &\dec
0.35 \\
FCC & Al &\dec 82.4 &\dec 34.6 &\dec 34.6 &\dec 34.6 &\dec 91.2
&\dec 0.32 \\
FCC & Al$_7$Li & \dec 75.6 &\dec 42.7 &\dec 42.7 &\dec 42.7 &\dec
107.7 &\dec 0.26 \\
L1$_2$ & Al$_3$Li &\dec 69.1 &\dec 45.3 &\dec 45.3 &\dec 45.3 &\dec
111.5 &\dec 0.23 \\
L1$_2$ & Al$_3$Li$^{\rm a}$ & & & & &\dec 91.0 & \\
\end{tabular}
\label{anitab}
\tablenotes{$^{\rm a}$Experiment, Ref.~\cite{Nobel82}, interpolated
to 25\% Lithium content.}
\end{table}

\end{document}